\documentclass[reqno,11pt]{amsart}

\IfFileExists{mymtpro2.sty}{%
  \usepackage[subscriptcorrection]{mymtpro2}
}{}

\usepackage{a4,amssymb}

\newtheorem{theorem}{Theorem}

\theoremstyle{remark}
\newtheorem{remark}[theorem]{Remark}
\newtheorem{myremarks}[theorem]{Remarks}

\newenvironment{acknowledgement}{\par\medskip\noindent\emph{Acknowledgements.}}

    {\end{nummer}\end{myremarks}}

\newcounter{numcount}
\newcommand{\labelnummer}{\mbox{\normalfont (\roman{numcount})}}%

\makeatletter

\newenvironment{nummer}%
  {\let\curlabelspeicher\@currentlabel%
    \begin{list}{\labelnummer}%
      {\usecounter{numcount}\leftmargin0pt%
        \topsep0.5ex\partopsep2ex\parsep0pt\itemsep0ex\@plus1\p@%
        \labelwidth2.5em\itemindent3.5em\labelsep1em%
      }%
    \let\saveitem\item%
    \def\item{\saveitem%
      \def\@currentlabel{{\upshape\curlabelspeicher}$\,$\labelnummer}}%
    \let\savelabel\label%
    \def\label##1{\savelabel{##1}%
      \@bsphack%
        \ifmmode\else%
          \protected@write\@auxout{}%
          {\string\newlabel{##1item}{{\labelnummer}{\thepage}}}%
        \fi%
      \@esphack%
    }%
  }{\end{list}}%

\renewcommand{\appendix}{\def\thesection{\textsc{Appendix}}}

 \let\leq\le
 \let\geq\ge

\DeclareMathOperator{\tr}{tr\kern1pt}

\makeatletter

\newif\ifper\pertrue
\def\per{.}

\def\bti{\@ifnextchar[\bbti\bbbti}
\def\bbti[#1]#2{#2, #1.}
\def\bbbti#1{#1.}

\def\z{\@ifnextchar[\zz\zzz}
\def\zz[#1]#2#3#4#5{\perfalse\emph{#2} \textbf{#3}, #4 (#5) [#1]}
\def\zzz#1#2#3#4{\emph{#1} \textbf{#2}, #3 (#4)\ifper\per\fi\pertrue}

\def\pub{\@ifstar\pubstar\pubnostar}
\def\pubnostar{\@ifnextchar[\@@pubnostar\@pubnostar}
\def\@@pubnostar[#1]#2#3#4{#2, #3, #4, #1\ifper\per\fi\pertrue}
\def\@pubnostar#1#2#3{#1, #2, #3\ifper\per\fi\pertrue}
\def\pubstar[#1]#2#3#4{\perfalse #2, #3, #4 [#1]\pertrue}

\makeatother

\newcommand{\beq}{\begin{equation}}
\newcommand{\eeq}{\end{equation}}
\newcommand{\ba}{\begin{array}}
\newcommand{\ea}{\end{array}}
\newcommand{\bea}{\begin{eqnarray}}
\newcommand{\eea}{\end{eqnarray}}

\newcommand{\R}{\mathbb{R}}

\newcommand{\Z}{\mathbb{Z}}

\def\P{I\kern-.30em{P}}
\def\E{I\kern-.30em{E}}
\renewcommand{\E}{\mathbb{E}\mkern2mu}
\renewcommand{\P}{\mathbb{P}}

\begin{document}

\title[Quantum transport for white-noise potentials]{Transport of a quantum particle in a time-dependent white-noise potential}

\author[P.\ D.\ Hislop]{Peter D.\ Hislop}
\address{Department of Mathematics,
    University of Kentucky,
    Lexington, Kentucky  40506-0027, USA}
\email{peter.hislop@uky.edu}

\author[K.\ Kirkpatrick]{Kay Kirkpatrick}
\address{Department of Mathematics \\
University of Illinois, Urbana-Champaign \\
Urbana, IL 61801}
\email{kkirkpat@illinois.edu}

\author[J.\ Schenker]{Jeffrey Schenker}
\address{Department of Mathematics \\
Michigan State University \\
East Lansing, MI}
\email{jeffrey@math.msu.edu}

\author[S.\ Olla]{Stefano Olla}
\address{Universit\'e Paris-Dauphine, PSL Research University \\
CNRS UMR CEREMADE \\
75016 Paris, France}
\email{olla@ceremade.dauphine.fr}


\begin{abstract}
We show that a quantum particle in $\R^d$, for $d \geq 1$, subject to a white-noise potential, moves superballistically in the sense that the mean square displacement $\int \|x\|^2 \langle \rho(x,x,t) \rangle ~dx$ grows like $t^{3}$ in any dimension. The white noise potential is Gaussian distributed with an arbitrary spatial correlation function and a delta correlation function in time. This is a known result in one dimension \cite{fischer-leschke-muller,jayannavar-kumar1}. The energy of the system is also shown to increase linearly in time. We also prove that for the same white-noise potential model on the lattice $\Z^d$, for $d \geq 1$, the mean square displacement is diffusive growing like $t^{1}$. This behavior on the lattice is consistent with the diffusive behavior observed for similar models in the lattice $\Z^d$ with a time-dependent Markovian potential \cite{ks}.
\end{abstract}

\maketitle \thispagestyle{empty}


\section{Statement of the Problem and Result}\label{sec:introduction}

A quantum particle in a random potential can move diffusively, ballistically, or superballistically depending on the circumstances.
In this note, we derive some results about the mean square displacement of a quantum particle subject to a time-dependent white noise Gaussian potential $V_\omega (x,t)$ that is correlated in space and uncorrelated in time. We prove that the mean square displacement is superballistic for models on $\R^d$ and diffusive for models on the lattice $\Z^d$.

We consider the Schr\"odinger equation with a time-dependent potential given by
\beq\label{eq:Schrod1}
i \hbar \partial _t \psi(x,t) = -\frac{\hbar^2}{2m} \Delta \psi(x,t) + V_\omega(x,t)\psi(x,t) ,
\eeq
where $x$ is in $\R^d$ (resp.\ $\Z^d$) and the operator $-\Delta$ is the Laplacian on $\R^d$ (resp.\ discrete Laplacian on $\Z^d$).
The potential $V_\omega(x,t)$ is a mean zero Gaussian stochastic process with covariance
\beq\label{eq:covariance}
\langle V_\omega(x,t) V_\omega(x',t') \rangle = V_0^2 g(x-x') \delta (t-t') ,
\eeq
where the strength of the disorder is $V_0 > 0$,
and the spatial correlation function $g \in C^2 (\R^d;\R)$ is a real, even function with sufficiently rapid decay at infinity. We assume the physically reasonable convexity condition that $|( \nabla g)(0)| = 0$ and that the Hessian matrix $Hess (g)(0)$ is negative definite.
The angular brackets in \eqref{eq:covariance} denote averaging with respect the Gaussian probability measure.

Generalizing from the single particle wave function $\psi$,
the evolution of a density matrix $\rho \in \mathcal{I}$, where $\mathcal{I}$ is the ideal of trace class operators,
is governed by the quantum stochastic Liouville equation
\beq\label{eq:evolution0}
i \hbar \frac{\partial \rho}{\partial t} = [ H_\omega (\lambda),  \rho],
\eeq
and an initial condition $\rho(t=0) = \rho_0 \in \mathcal{I}$, a non-negative trace-class operator.
We assume that the density matrix $\rho \in \mathcal{I}$ has a kernel $\rho (x',x, t)$.
It follows from \eqref{eq:evolution0} that the kernel
satisfies the quantum stochastic Liouville equation:
\beq\label{eq:evolution1}
i \hbar \frac{\partial \rho}{\partial t}(x,x',t) = L \rho (x,x',t),
\eeq
with initial condition $\rho (x,x',0)$
corresponding to $\rho(t=0) = \rho_0 \in \mathcal{I}$.
The generator of the time evolution $L$, called the Liouvillian and appearing in \eqref{eq:evolution1}, is given by:
\beq\label{eq:generator1}
(Lf)(x,x') = \frac{\hbar^2}{2m} [ (\Delta_x f)(x,x') - (\Delta_{x'} f)(x,x')] + [ V_\omega (x,t) - V_\omega(x',t)]f(x,x') .
\eeq
For example, if
$\rho$ is a pure state density matrix $\rho = P_\psi$,
where $P_\psi$ projects onto the normalized state $\psi$,
the kernel $\rho (x,x', t)  = \overline{\psi}(x,t) \psi(x',t)$.

In general, we will consider kernels $\rho(x,x',t)$ of density matrices $\rho$ solving the stochastic Liouville equation \eqref{eq:evolution1} with initial kernels $\rho_0 (x,x')$, corresponding to a non-negative trace-class operators $\rho_0$. We assume that $\rho_0$ has a well-behaved kernel so that $\langle \|x\|^2 \rangle_{\rho_0} < \infty$. The solution $\rho_t$ has a well-behaved kernel $\rho(x,x',t)$ so that ${\rm Tr} \rho_t = \int_{\R^d} ~\rho(x,x,t) ~dx < \infty$, for all $t \in \R$. We set the disorder $\lambda = 1$.
We denote by $\langle A \rangle_{\rho(t)} := \langle {\rm Tr} \{ A \rho (t) \} \rangle$ the average of the expectation of an observable $A$ in the state $\rho(t)$.

Equations \eqref{eq:Schrod1} and \eqref{eq:evolution1} are stochastic partial differential equations. The product of the white noise potential and the solution is interpreted in the Stratonovich sense. The details of the derivation of the partial differential equation for the averaged kernel $\langle \rho(x,x',t) \rangle$  of the density matrix \eqref{eq:schr4} are presented in the appendices,  sections \ref{sec:sde1} and \ref{sec:novikov1}.

\begin{theorem}\label{thm:main1}
We consider the stochastic Liouville equation \eqref{eq:evolution1} with Liouvillian $L$ given in \eqref{eq:generator1} with a white noise random potential with covariance given by \eqref{eq:covariance} satisfying [H1]--[H3] in section \ref{sec:laplace1}. Let $ \langle \rho (x,x',t) \rangle$ be the disorder-averaged density matrix, solving the evolution equation \eqref{eq:schr4} derived in section \ref{sec:laplace1}, with initial condition $\rho_0 (x,x')$.
\begin{enumerate}
\item \textbf{Continuum}: For $\R^d$, the disorder-averaged mean square displacement is {superballistic}, $\langle \|x\|^2 \rangle_{\rho (t)} \sim t^{3}$, that is,
$$
\langle \|x^2 \| \rangle_{\rho(t)} = \int_{\R^d} \|x\|^2 \langle \rho(x,x,t) \rangle ~d^d x  = B(V_0) t^3 + \mathcal{O}(t^2),
$$
in any dimension $d \geq 1$, where $B(V_0) = - \frac{1}{3 (2^d)} \left( \frac{V_0}{m} \right)^2 (\Delta g)(0) {\rm Tr} ( \rho_0 )$.

\item \textbf{Lattice}: For $\Z^d$, superballistic motion is suppressed and the disorder-averaged mean square displacement is {diffusive}, $\langle \|x(t)\|^2 \rangle_\rho \sim t$, that is,
$$
\langle \|x^2 \| \rangle_{\rho(t)}  = \sum_{x \in \Z^d} \|x\|^2 \langle \rho(x,x,t) \rangle  = D(V_0) t + \mathcal{O}(1),
$$
in any dimension $d \geq 1$. The diffusion constant $D = D(V_0)$ is proportional to $\sum_{m=1}^d \{(V_0/\hbar)^2 [g(0) - g(\hat{e}_m)]\}^{-1} {\rm Tr} ( \rho_0 )$ and strictly positive provided $V_0 > 0$ and $g(0) \neq g(\hat{e}_m)$, for $m=1, \ldots, d$, where $\{ \hat{e}_m \}$ is the standard orthonormal basis of $\Z^d$.

\vspace{.1in}
\end{enumerate}
If $V_0 = 0$ in either case, then the mean square displacement is $\mathcal{O}(t^2)$, so the motion is ballistic.
\end{theorem}


Theorem \ref{thm:main1} indicates that the type of the underlying space, whether it is the lattice $\Z^d$ or continuum $\R^d$, affects the quantum motion dramatically. In the absence of a potential, $V_0 = 0$ in \eqref{eq:Schrod1}, it is known that a free quantum particle on $\Z^d$ or $\R^d$ moves \emph{ballistically}.
In contrast, on $\Z^d$, with appropriate \emph{bounded} noise $V_\omega$ and $V_0 \neq 0$, a quantum particle travels \emph{diffusively} \cite{drf,fs,ks}. As will be discussed below, the difference in these two settings is tied to the unboundedness or boundedness of the unperturbed operator $H_0 = - \Delta$.

In classical mechanics, a particle moving in the continuum under the influence of suitable time-independent random potentials behaves diffusively like a Brownian motion so that the mean square displacement proportional to $t$. By contrast, a classical particle moving on a lattice in a time-independent random potential has mean square displacement bounded by a constant \cite{md}. The classical analog of part (1) of Theorem \ref{thm:main1} is a classical particle subject to a force described by a time-dependent white-noise Gaussian potential $V_\omega$ satisfying \eqref{eq:covariance1}. The particle position's $q(t) \in \R^d$ satisfies the equation $q''(t) = - \nabla V_\omega(q(t))$. Let us assume that the covariance function $g$ is smooth. Upon integrating the equation of motion for the velocity $q'(t)$, squaring, and taking the average using the relation \eqref{eq:covariance1},
we heuristically find that $\langle \| q'(t)\|^2 \rangle = \frac{1}{2} V_0^2 (\Delta g)(0) t$.
This suggests that $q'(t) \sim t^{\frac{1}{2}}$, so that
$q(t) \sim t^{\frac{3}{2}}$, similar to, and consistent with, the quantum motion described in part (1) of Theorem \ref{thm:main1}.



In order to put our results in context,
we recall the localization-delocalization problem for a random Schr\"odinger operator with a static random potential $V_\omega(x,t) = V_\omega(x)$. Anderson localization is known to occur in one-dimension or for a strong static random potential in higher dimensions analogous to the recurrence of random walks in one- and two-dimensions, and transience in three-dimensions, For random Schr\"odinger operators, large disorder can cause recurrence leading to Anderson localization \cite{AM,ASFH,Anderson}. By contrast, a main open problem is to prove that there is quantum diffusion at weak disorder in dimensions $d \geq 3$ for high energies for the continuum models, or for energies near the center of the band for lattice models.
This type of static disorder allows correlations between past and present when the particle revisits any part of the environment where it has been before, and thus is hard to handle in the delocalization regime at weak disorder, especially combined with recurrence of random walks in one- and two-dimensions, and transience in three-dimensions. Instead, we consider a time-dependent random potential $V_\omega(x,t)$ that can reduce or remove such temporal correlations. This allows more results to be proved, such as superballistic behavior for models on $\R^d$ and diffusive behavior on $\Z^d$, at all energies, while still providing an interesting and physically meaningful setting. This is similar to questions about random walks in random environments, where a static random environment is more challenging to understand, and a dynamic random environment can be more tractable.


\subsection{Related results}\label{subsec:relatedResults1}

Fischer, Leschke, and M\"uller \cite{fischer-leschke-muller, fischer-leschke-muller2} proved a result similar to part (1) of Theorem \ref{thm:main1}. They work in the Weyl-Wigner-Moyal formulation of quantum mechanics on phase space $\R^d \times \R^d$. Their Hamiltonians have the form $H(p,q) + N_t (p,q)$ where the white-noise potential $N_t(p,q)$ satisfies a covariance relation similar to \eqref{eq:covariance1} where the covariance function $g$ depends on $p$ as well as $q$. For the case $H(p,q) = p^2 / (2m)$, they prove that $q(t) \sim t^{\frac{3}{2}}$, see \cite[(2.60e)]{fischer-leschke-muller}.
We remark that in a related paper, Fischer, Leschke, and M\"uller \cite{fischer-leschke-muller} state
if one adds an external magnetic field to the model on $\R^2$, then
the superballistic motion is suppressed and one recovers diffusive motion. This is proved
by M\"uller \cite[Beispiel 2.36]{muller1} in his thesis.

Concerning lattice models, Ovchinnikov and Erikhman \cite{ovchinnikov-erikhman1} proved a result similar to part (2) of Theorem \ref{thm:main1} for one-dimensional linear random chains using a Laplace transform method similar to that used later in \cite{jayannavar-kumar1}. We discuss their work further at the end of section \ref{sec:lattice1}.


Our methods are very different from those of \cite{fischer-leschke-muller, fischer-leschke-muller2} and generalize many results of \cite{ovchinnikov-erikhman1} to all dimensions. Furthermore, our methods are applicable to both the continuum and lattice models.




There has been some discussion in the physics literature concerning models for which the delta correlation in time in \eqref{eq:covariance} is replaced by a more general function $h(t - t')$, so-called colored noise (see sections \ref{sec:sde1} and \ref{sec:novikov1}). Golubovi\'c, Feng, and Zeng \cite{gfz1} studied Gaussian random variables with a covariance
\beq\label{eq:covariance2}
\langle V_\omega(x,t) V_\omega(x',t') \rangle = V_0^2 e^{- \frac{\|x-x'\|^2}{\ell^2} - \frac{(t-t')^2}{\tau^2}} ,
\eeq
where $\ell$ and $\tau$ are effective spatial and temporal correlation lengths.
Beginning with the Schr\"odinger equation on $\R^d$, they derive an effective Fokker-Planck equation for the velocity distribution. Their main result is that temporal correlations cause superballistic behavior but with different exponents than in Theorem \ref{thm:main1}. Golubovi\'c, Feng, and Zeng \cite{gfz1} showed that the mean square displacement is superballistic: for $d=1$ is $t^{\frac{12}{5}}$, while for $d > 1$, it is $t^{\frac{9}{4}}$.
Rosenbluth \cite{rosenbluth}, disagreeing with the derivation of the Fokker-Planck equation
in \cite{gfz1}, derived another equation and used it to show that the mean square displacement for $d=1$ is $t^{\frac{12}{5}}$, while for $d > 1$, it is $t^{2}$, ballistic motion.

The subject of stochastic acceleration for classical systems has been frequently discussed in the literature. For example, in the work of Aguer, et.\ al. \cite{aguerDebievreLafitteParris}, the authors present theoretical and numerical results that indicate, amongst other results, that the mean squared displacement in a space and time homogeneous random field with rapid decay in space but not necessarily in time, for which the force field is not a gradient field, is superballistic with $\langle \| q (t) \|^2 \rangle \sim t^{\frac{8}{3}}$, for dimensions $d \geq 1$. When the force field is a gradient field, the motion is superballistic only in dimension one, and ballistic for $d \geq 2$. The models include the inelastic, nondissipative soft Lorentz gas. In Soret and De Bi\`evre \cite{soretDebievre1}, the authors treat a simplified random, inelastic, Lorentz gas model and prove, rougly speaking, that the average velocity $\langle \| q'(t) \| \rangle \sim t^{\frac{1}{5}}$, for dimensions $d \geq 5$.


\subsection{Contents of the paper}\label{contents1}

        In section \ref{sec:laplace1}, we prove superballistic motion for the model on $\R^d$ extending the  Laplace transform method introduced by Jayannavar and Kumar \cite{jayannavar-kumar1} for one-dimensional models. After reviewing this approach, we extend it to multi-dimensional models using the method of characteristics.
We then discuss in section \ref{sec:lattice1} the slowing effect of the lattice $\Z^d$. The proof of diffusive motion on the lattice requires modification of the Laplace transform technique. We conclude with two appendices. In the first, section \ref{sec:sde1}, we
discuss the Stratonovich interpretation of the stochastic differential equation and in the second, section \ref{sec:novikov1}, we present a calculation on Gaussian correlations.

%
%
%
%
%
%
%


\section{Laplace transform approach to quantum motion on $\R^d$}\label{sec:laplace1}

Jayannavar and Kumar \cite{jayannavar-kumar1} studied the evolution of a pure state density matrix in one dimension. They chose a Gaussian initial state $\psi(x,0) = C_\sigma e^{- x^2/ 4 \sigma^2}$ so that $\rho_0(x,x^\prime) = \psi(x^\prime,0) \psi(x,0)$.
The solution $\rho(x^\prime, x, t)$ satisfies the stochastic Liouville equation \eqref{eq:evolution1}-\eqref{eq:generator1}. Averaging over the Gaussian disorder, Jayannavar and Kumar \cite{jayannavar-kumar1} solve the resulting equation for the averaged density kernel $\langle \rho(x^\prime, x, t) \rangle$. We first review the construction of the solution in one-dimension. After a review of the method of characteristics in section \ref{subsubsec:characteristics1}, we prove an explicit solution for the averaged density  kernel in any dimension for any initial density matrix, not necessarily a pure state.
We prove that the quantum motion is like $\langle \| x(t) \| \rangle \sim t^{3/2}$ in any dimension $d$ on $\R^d$. This exponent $3/2$ is independent of the dimension and appears to be independent of the type of disorder.

\begin{description}
\item[{[H1]}] The  potential $V_\omega(x,t)$ is a mean zero Gaussian stochastic
 process $\langle V(x,t) \rangle = 0$ with covariance
\beq\label{eq:covariance1}
\langle V_\omega(x,t) V_\omega(x',t') \rangle = V_0^2 g(x-x') \delta (t-t') .
\eeq

\item[{[H2]}] The strength of the potential is $V_0 > 0$.

\item[{[H3]}] The spatial correlation function $g \in C^2 (\R^d, \R)$ is assumed to be a real, {\it even} function with sufficiently rapid decay at infinity. It satisfies the physically reasonable convexity condition that $|( \nabla g)(0)| = 0$ and that $Hess (g)(0)$ is negative definite.
\end{description}

By sufficiently rapid decay in [H3], we need that, at a minimum, $g$ and its first and second partial derivatives belong to $L^1 (\R^d)$. 
Since a correlation function must be the Fourier transform of a positive measure, the reality of $g$ forces it to be an even function. A common example of a correlation function $g$ satisfying [H3] is a Gaussian function $g(x) = e^{- \sum_{i,j=1}^d A_{ij} x_i x_j}$, for a positive definite matrix $A$.

For a random variable $K$, we denote by $\langle K \rangle$ the expectation with respect to the Gaussian process.

Since the Schr\"odinger equation \eqref{eq:Schrod1} and the Liouville equation \eqref{eq:evolution1}-\eqref{eq:generator1} with a white noise potential \eqref{eq:covariance1} are stochastic partial differential equations, the singular term $V_\omega (x,t) \psi(x,t)$ requires some interpretation.
The Gaussian nature of the process, however, allows us to easily derive a partial differential equation for the
averaged density $\rho$.

The basic approach of Jayannavar and Kumar \cite{jayannavar-kumar1} is as follows.
If $\psi(x,t)$ satisfies the Schr\"odinger equation
\beq\label{eq:schr1}
i\hbar \partial_t \psi = H \psi, ~~H = -\frac{\hbar^2}{2m} \Delta + V_\omega(\cdot,t),
\eeq
then the pure state density matrix $\rho(x',x,t)$, given by $\rho(x',x,t) = \overline{\psi}(x',t) \psi(x,t),$
satisfies the equation
\bea\label{eq:schr2}
\partial_t \rho(x',x,t)  & =  & - \frac{i\hbar}{2m} ( \Delta_{x'} - \Delta_{x}) \rho(x',x,t)  \nonumber \\
 & & - \frac{i}{\hbar}( V_\omega(x,t) - V_\omega(x',t)) \rho(x',x,t) \nonumber \\
 &= & - \frac{i}{\hbar} L \rho (x',x,t),
 \eea
the Liouvillian equation \eqref{eq:evolution1}-\eqref{eq:generator1}.
In general, we now assume that $\rho(x',x,t)$ is the kernel of a density matrix that solves the Liouville equation \eqref{eq:evolution0} so that the kernel satisfies \eqref{eq:schr2}
with an initial condition $\rho_0 (x',x)$. We normalize the initial conditions so ${\rm Tr} \rho_0 = \int_{\R^d} ~
 \rho_0 (x,x ) ~dx = 1$. We will always assume that the diagonal of the kernel of the initial density matrix $\rho_0$ satisfies
 $$
 \int_{\R^d} ~ \| x \|^2 \rho_0 (x,x) ~dx < \infty,
 $$
 in the continuum case, or, for the $\Z^d$ case,
 $$
 \sum_{x \in \Z^d} ~ \| x \|^2 \rho_0 (x,x) < \infty,
 $$


Taking the expectation of
\eqref{eq:schr2}, we obtain
\bea\label{eq:schr3}
\partial_t \langle \rho(x',x,t) \rangle  & =  & - \frac{i\hbar}{2m} ( \Delta_{x'} - \Delta_{x}) \langle \rho(x',x,t) \rangle  \nonumber \\
 & & - \frac{i}{\hbar}\langle ( V_\omega(x',t) - V_\omega(x,t)) \rho(x',x,t) \rangle .
 \eea

We next use Novikov's Theorem \cite{novikov1} to compute the average $\langle V_\omega(z,t) \rho(x',x,t) \rangle$, where $z$ denotes
  $x$ or $x'$, with respect to the Gaussian random variables under the covariance assumption \eqref{eq:covariance1}.
From section \ref{sec:novikov1}, we obtain
\beq\label{eq:averaging-result1}
\langle V_\omega(z,t) \rho(x',x,t) \rangle = \frac{i V_0^2}{2 \hbar} [ g(z-x') - g (z-x)] \langle \rho(x', x, t)
\rangle .
\eeq
Using this result, we find that \eqref{eq:schr3} may be written as
\bea\label{eq:schr4}
\partial_t \langle \rho(x',x,t) \rangle & =  & - \frac{i\hbar}{2m} ( \Delta_x - \Delta_{x'}) \langle \rho(x',x,t) \rangle \nonumber \\
 & & - \left(\frac{V_0}{\hbar} \right)^2 [ g(0) - g(x-x')] \langle \rho(x',x,t) \rangle .
\eea


We introduce new variables $X := x+x'$ and $Y := x- x'$ into the equation \eqref{eq:schr2}.
We write $R(X,Y,t)$ for $\langle \rho(x',x,t) \rangle$.
The Laplace transform of $R(X,Y,t)$
with respect to $t$ is
\beq\label{eq:lt0}
\widetilde{R}(X,Y,s) = \int_0^\infty e^{-st} R(X,Y,t) ~dt.
\eeq
Using the initial condition $R(X,Y,0) := \langle \rho (x',x,0) \rangle$, and defining
\beq\label{eq:hYs}
h(Y,s) :=  s + \left(\frac{V_0}{\hbar } \right)^2 [ g(0) - g(Y)] ,
\eeq
 we obtain
\beq\label{eq:lt9}
\frac{2i\hbar}{m} \nabla_X \cdot \nabla_Y {\widetilde{R}}(X,Y,s) + h(Y,s) {\widetilde{R}}(X,Y,s)
  =   {R}(X,Y,0)
  \eeq

We define the Fourier transform of the density matrix with respect to $X$:
\beq\label{eq:fourier-trans1}
\widehat{R}(k,Y,t) := \int_{\R^d} e^{i k\cdot X} R(X,Y,t) ~dX .
\eeq
Taking the Fourier transform of the equation \eqref{eq:lt9} with respect to $X$, we obtain
\beq\label{eq:lt10}
k \cdot \nabla_Y \widehat{\widetilde{R}}(k,Y,s) + \frac{m}{2\hbar}  h(Y,s) \widehat{\widetilde{R}}(k,Y,s)
  =   \frac{m}{2\hbar} \widehat{{R}}(k,Y,0)
  \eeq
We will solve this equation for $\widehat{\widetilde{R}}(k,Y,s)$ in one dimension in section
\ref{subsec:one-dim1} and in any dimension in section \ref{subsec:multidim1}. Setting $Y=0$, this yields the Laplace transform of the function ${\widetilde{R}}(k,0,t)$.
Combining \eqref{eq:fourier-trans1} and the fact that $R(X,0,t) = \langle \rho(x,x,t)$,
we find that the second moments of the position vector $x$
are given by
\beq\label{eq:second-moments1}
\int_{\R^d} ~x_i x_j \langle \rho(x,x,t) \rangle ~d x =  - \frac{1}{2^{d+2}} \frac{\partial^2}{\partial k_i \partial k_j} \left. {\widehat{R}}(k,0,t) \right|_{k=0}.
\eeq
The mean square displacement is given by
\bea\label{eq:second-moments2}
\langle \|x\|^2 \rangle_{\rho (t)}  &=& \int_{\R^d} ~\|x\|^2 \langle \rho(x,x,t) \rangle ~dx \nonumber \\
 &=& - \left( \frac{1}{2^{d+2}} \right) \Delta_k \left. {\widehat{R}}(k,0,t) \right|_{k=0}.
\eea

In section \ref{sec:lattice1}, we will prove that quantum transport with the white noise potential
on $\Z^d$ is diffusive.
The diffusion coefficient matrix $[d_{ij}]$ is defined by
\beq\label{eq:diff1}
d_{ij} =: \lim_{t \rightarrow \infty} - \frac{1}{t} \frac{1}{2^{d+2}} \frac{\partial^2}{\partial k_i \partial k_j} \left. {\widehat{R}}(k,0,t) \right|_{k=0},
\eeq
and the diffusion constant $D$ is the trace of this matrix
\beq\label{eq:diff2}
D = {\rm Tr} [ d_{ij} ] = \sum_{i=1}^d d_{ii} .
\eeq



\subsection{Solution of the one-dimensional problem}\label{subsec:one-dim1}

In the one-dimensional case, equation \eqref{eq:lt9} for $\widetilde{R}(X,Y,s)$ agrees with \cite[eqn. 10]{jayannavar-kumar1}. We now take the Fourier transform with respect to $X$. This results in the ordinary differential equation:
\beq\label{eq:one-dim-lt1}
\frac{d}{dY} \widehat{\widetilde{R}}(k,Y,s) + \frac{m}{2\hbar k}  h(Y,s) \widehat{\widetilde{R}}(k,Y,s)
  =   \frac{m}{2\hbar k} \widehat{{R}}(k,Y,0)
  \eeq
We integrate this equation with the boundary condition $\widehat{\widetilde{R}}(k,Y=b,s) = 0$ and obtain
  \beq\label{eq:one-dim-lt2}
 \widehat{\widetilde{R}}(k,Y,s) = \frac{m}{2\hbar k} \int_b^Y e^{- \int_z^Y  \frac{m}{2\hbar k}  h(w,s) dw } \widehat{R}(k,z,0) ~dz.
 \eeq
 Taking $b = - \infty$ imposes the physically reasonable boundary condition $\lim_{Y \rightarrow - \infty} \widehat{\widetilde{R}}(k,Y,s) =0$ corresponding to the decay of the kernel of the density matrix. We now take $Y=0$. Using the definition of $h$ in \eqref{eq:hYs} and changing variables with $\tilde{z} := - (m / (2\hbar k))z$, we obtain
 \beq\label{eq:one-dim-lt3}
 \widehat{\widetilde{R}}(k,0,s) = \int_0^\infty e^{- sz} e^{-\left( \frac{V_0}{\hbar} \right)^2[ g(0)z -  \int_0^z g \left(\frac{-2\hbar k}{m}w \right) dw]}
       ~\widehat{R}(k,-2\hbar kz/m,0) ~dz.
 \eeq
The integral in \eqref{eq:one-dim-lt3} is in the form of a Laplace transform, so
\beq\label{eq:one-dim-lt4}
 \widehat{{R}}(k,0,t) =  e^{-\left( \frac{V_0}{\hbar } \right)^2[ g(0)t -  \int_0^t g \left(\frac{-2\hbar k}{m}w \right) dw]}
       ~\widehat{R}(k,-2\hbar kt/m,0).
 \eeq
Note that one does not have to specify the spatial correlation function $g$ nor the initial density matrix $\rho_0$.

As follows from \eqref{eq:second-moments2}, the second moment of the position operator
is calculated from two derivatives with respect to $k$ of $\widehat{R}$:
\beq\label{eq:moment1}
\langle x(t)^2 \rangle_\rho = - \frac{1}{8} \left. \frac{d^2}{dk^2} \widehat{R}(k,0, t) \right|_{k=0} .
\eeq
We define the phase function in \eqref{eq:one-dim-lt4} to be
\beq\label{eq:phase1}
\Phi(k,t) := -\left( \frac{V_0}{\hbar } \right)^2[ g(0)t -  \int_0^t g \left(\frac{-2\hbar k}{m}w \right) dw],
\eeq
and note that $\Phi (0,t) = \Phi^\prime (0,t) = 0$ due to [H3].
In terms of the phase function, we have
$$
 \widehat{{R}}(k,0,t) := e^{\Phi(k,t)} \widehat{R}(k,-2\hbar kt/m,0) .
 $$
The computation of the second derivative yields
 \bea\label{eq:moment3}
 -4 \frac{d^2}{dk^2} \widehat{R}(k,0, t) & = & \left[ \frac{d^2 \Phi(k,t)}{dk^2} + \left( \frac{d \Phi(k,t)}{dk} \right)^2 \right]  \widehat{R}(k,-2\hbar kt/m , t) e^{\Phi(k,t)}
    \nonumber \\
     & & +  \left[ 2 \left( \frac{d \Phi(k,t)}{dk} \right) \left( \frac{d }{dk} \widehat{R}(k,-2\hbar kt/m,0)  \right)  \right. \nonumber \\
      & & + \left. \frac{d^2}{dk^2} \widehat{R}(k,-2\hbar kt/m,0)      \right] e^{\Phi(k,t)} . \nonumber \\
      & &
\eea

The crucial part of the calculation that gives the leading behavior in $t$
is the second derivative of the phase function in \eqref{eq:phase1}:
\bea\label{eq:moment2}
\left. \frac{d^2 \Phi(k,t)}{dk^2} \right|_{k=0} & = & \left. \left( \frac{V_0}{\hbar} \right)^2 \int_0^t g'' (  -2 \hbar kw / m) \left( \frac{2\hbar w}{m} \right)^2 ~dw \right|_{k=0} \nonumber \\
 &=& \frac{1}{3} \left( \frac{2V_0}{m} \right)^2   g'' (0) t^3 ,
\eea
where integration over $w \in [0, t]$ gives the $t^3$ term. This shows that
\beq\label{eq:moment4}
\langle  x^2 \rangle_{\rho (t)} = - \frac{1}{6} \left( \frac{V_0}{m} \right)^2 g''(0) [ {\rm Tr} \rho_0 ]  t^3
 + \mathcal{O}(t^2).
\eeq
We note that the evenness of $g$ is important here: If $g'(0) \neq 0$ then the term
$$
\left( \frac{d \Phi(k,t)}{dk} \right)^2
$$
is $g'(0)^2 t^4$ and dominates the behavior of the second moment.

If the random potential vanishes, the correlation function $g=0$ and the phase $\Phi = 0$. We see that the last term on the right of \eqref{eq:moment3} behaves like $t^2$ so the motion is ballistic.

\subsection{The multidimensional continuum problem}\label{subsec:multidim1}

A similar approach may be taken in order to compute the mean square displacement on $\R^d$ in any dimension. The additional component required for this is the solution of a nonhomogeneous transport equation.

\subsubsection{The method of characteristics}\label{subsubsec:characteristics1}

In this section, we review the method of characteristics for a semilinear transport equation:
\beq\label{eq:semilin1}
k \cdot \nabla_x u(x) = c(x,u).
\eeq
In our case $c(x,u) = - h(x)u(x) + m(x)$.
The characteristic equations for $(x(s), z(s))$ are:
\beq\label{eq:charact1}
\frac{dx}{ds} = k,
\eeq
and
\beq\label{eq:charact2}
\frac{dz}{ds} = c(x,z) = - h(x) z + m(x) .
\eeq
The first equation integrates $x(s) = ks + k_0$, with $k, k_0 \in \R^d$, and $s \in \R$.
The solution of the first-order ordinary differential equation \eqref{eq:charact2} for $z(s)$,
\beq\label{eq:ode1}
\frac{dz}{ds} + h(x(s))z(s) = m(x(s)) ,
\eeq
has the form
\beq\label{eq:ode2}
z(s) = \int_b^s e^{- \int_w^s h(x(w')) ~dw'} m(x(w)) ~dw,
\eeq
with the boundary condition $z(b) = 0$ at $b$ that is determined by the problem.
We recall that $z(s) = u(x(s))$ solves the original equation \eqref{eq:semilin1}.


\subsubsection{The general solution}\label{subsubsec:multidim1}

We apply the method of characteristics to the semilinear transport equation \eqref{eq:lt10}. We let $(Y(v), z(v))$ be the solutions of the corresponding characteristic equations \eqref{eq:charact1} and \eqref{eq:charact2} so that $z(v) = \widehat{\widetilde{R}}(k,Y(v),s)$. The function $Y(v) = kv + k_0$, for $w \in \R$ and $k, k_0 \in \R^d$, solves the characteristic equation \eqref{eq:charact1}. The function $z(v)$ solves the second equation \eqref{eq:ode1}: \beq\label{eq:pde1}
  \frac{d}{dv} \widehat{\widetilde{R}}(k,Y(v),s) + h(Y(v),s) \widehat{\widetilde{R}}(k,Y(v),s)
  =   \frac{m}{2\hbar } \widehat{{R}}(k,Y(v),0),
  \eeq
where as before in \eqref{eq:hYs} we define
$$
h(Y(v),s) :=  s + \left(\frac{V_0}{\hbar } \right)^2 [ g(0) - g(Y(v))]  .
$$
As in \eqref{eq:ode2}, the solution is
\beq\label{eq:pde2}
\widehat{\widetilde{R}}(k,Y(v),s) =
\frac{m}{2\hbar } \int_{-\infty}^v e^{- \int_w^v \frac{m}{2\hbar } h(Y(w'),s) ~dw'} \widehat{{R}}(k,Y(w),0)  ~dw ,
\eeq
with the boundary condition $\widehat{\widetilde{R}}(k,Y(-\infty),s) = 0$ as described below \eqref{eq:one-dim-lt2}.

Following the reductions used in the one-dimensional case, we re-express the integral in \eqref{eq:pde2} as a Laplace transform. We finally obtain
\beq\label{eq:ft10}
\widehat{R}(k, Y(0),t) = \widehat{R}\left(k, Y\left(-2\hbar t/m\right), 0\right) e^{- \left( \frac{V_0}{\hbar }\right)^2 \left[ g(0)t - \int_0^t g\left(Y\left(\frac{-2\hbar s}{m}\right)\right)~ds \right] }.
\eeq

We now compute the second moment of the position operator following \eqref{eq:second-moments2}. The integration constant $k_0$, appearing in the solution of the characteristic equation \eqref{eq:charact1}, is set equal to zero.
As in \eqref{eq:phase1}, we define the phase function $\Phi (k,t)$ as
\beq\label{eq:phase2}
\Phi (k,t) = - \left( \frac{V_0}{\hbar} \right)^2 \left[ g(0) t - \int_0^{t} g\left( - \frac{2 \hbar s k}{m} \right) ~ds \right] .
\eeq
In analogy with \eqref{eq:one-dim-lt4}, the crucial
computation of the derivative with respect to $k$ relies on the fact that $Y(v) = vk$, $v \in \R$. Consequently, we have a term identical
to \eqref{eq:moment2} which gives the superballistic behavior \eqref{eq:moment3}. In analogy with the calculation
\eqref{eq:moment3}, we find
\beq\label{eq:second-deriv-nonhomog1}
\left. \frac{\partial^2 \Phi}{\partial k_j \partial k_i} (k,t)\right|_{k=0} =  \frac{1}{3} \left( \frac{2 V_0}{m} \right)^2 ~ \frac{\partial^2 g}{\partial k_j \partial k_i}(k_0) t^3.
\eeq
In light of \eqref{eq:second-moments2}, this yields:
\bea\label{eq:moment1cont}
\langle  \| x \|^2 \rangle_{\rho (t)} & =  & - \frac{1}{2^{d+2}} \sum_{j=1}^d \partial_j^2 \hat{R}(k,0,t) |_{k=0} \nonumber \\
 & = & - \frac{1}{3} \frac{1}{2^d} \left( \frac{ V_0}{m} \right)^2 ( \Delta g)(0) ~ {\rm Tr} ( \rho_0 ) ~  t^3 + \mathcal{O}(t^2).
 \eea

\begin{remark}
As is shown in \cite{fischer-leschke-muller, fischer-leschke-muller2}, the kinetic energy $H_0 = - \Delta$ in this situation satisfies a linear in time lower bound in the state $\rho$:
$$
\langle H_0 \rangle_{\rho (t)} \geq C_0 t, ~~~t > 0.
$$
This indicates that linear energy growth due to the unbounded white noise potential. Unlike the lattice case, the  kinetic energy operator $H_0$ is unbounded so an infinite amount of energy can be added to the system.
In \cite[section 3]{fischer-leschke-muller2}, the authors consider a model with white noise and dissipation through a linear coupling to a heat bath. They prove that the averaged energy remains bounded in certain situations.
\end{remark}

\section{The Laplace transform approach to quantum motion on $\Z^d$}\label{sec:lattice1}

The motion of a quantum particle restricted to a square lattice $\Z^d$ differs from the results in section \ref{sec:laplace1} primarily due to the fact that the lattice Laplacian is a bounded operator.
The Gaussian white-noise potential results in diffusive motion on the lattice.
This is in keeping with the result of Kang and Schenker \cite{ks}. They studied a similar problem on $\Z^d$ for which the potential $V_\omega(x,t)$ is a Markovian potential. They proved that the motion is diffusive.
We now turn to the proof of the second part of Theorem \ref{thm:main1}.



To prove part 2 of Theorem \ref{thm:main1}, we will pursue the same basic strategy as in section \ref{sec:laplace1} except an exact solution for the Laplace transform
as in \eqref{eq:one-dim-lt3} is no longer possible. Instead, we will prove that
\beq\label{eq:diffusion-condition1}
\int_0^\infty ~e^{-ts} \left( \sum_{x \in \Z^d} \| x \|^2 \langle \rho (x,x,t) \rangle  \right)  ~dt = \left. - \sum_{j=1}^d  \frac{\partial^2}{\partial k_j^2} \widehat{\widetilde{R}}(k,0,s) \right|_{k=0} = \frac{C_0}{s^2} + \mathcal{O} \left( \frac{1}{s} \right),
\eeq
for $C_0 > 0$ depending on $V_0$ and $g$.
Condition \eqref{eq:diffusion-condition1} is equivalent to $\langle \|x \|^2 \rangle_{\rho_t} \sim t$ indicating diffusive motion. We will use below the fact that the expressions in \eqref{eq:diffusion-condition1} are real.

Let $\{ \hat{e}_j \}$ be the standard orthonormal basis of $\Z^d$. The discrete Laplacian acting on a function $f$ at site $x \in \Z^d$ sums $f$ over all $2d$ nearest neighbors to $x$. We write $\{ \hat{f}_j \}_{j=1}^{2d}$ for the $2d$ nearest neighbor directions at the origin so that $\hat{f}_j = \hat{e}_j$, for $j=1, \ldots d$ and $\hat{f}_{d+j} = - \hat{e}_j$, for $j=1, \ldots, d$. With this notation, the discrete Laplacian $\Delta$ on $\ell^2(\Z^d)$ is the
finite-difference operator
\beq\label{eq:discrete-lap1}
( \Delta f)(x) = \sum_{y: |x-y| =1} ~f(y) = \sum_{j=1}^{2d} ~f(x + \hat{f}_j ) = \sum_{j=1}^{d} \left[ ~f(x + \hat{e}_j ) + f(x - \hat{e}_j) \right] .
\eeq
The Laplacian is normalized so its spectrum is $[-2d, 2d]$.
The Laplacian may be factored via directional derivatives $\nabla_i^\pm$ defined by
\beq\label{eq:discrete-deriv1}
( \nabla_i^\pm f)(x) = f(x \pm \hat{e}_i) - f(x) .
\eeq
These two finite-difference operators commute. The adjoint of $\nabla_i^\pm$ is $\nabla_i^\mp$.
In terms of these, the discrete Laplacian \eqref{eq:discrete-lap1} may be written as
\beq\label{eq:discrete-factor1}
(\Delta f)(x) = - \sum_{j=1}^d ~(\nabla_j^+ \nabla_j^- f)(x) + 2 d f(x)
\eeq

It is convenient to introduce new variables $X = x + x'$ and $Y = x - x'$.
In terms of these variables,
we obtain
\beq\label{eq:discrete-diff1}
\Delta_x - \Delta_{x'} = 2 \sum_{j=1}^{d} \left[ \nabla_{Y,j}^+ \nabla_{X,j}^-
+ \nabla_{X,j}^+ \nabla_{Y,j}^- \right] .
\eeq
As above, we write $R(X,Y,t)$ for the averaged density $\langle \rho (x',x,t)\rangle$. We then obtain from the fundamental equation \eqref{eq:schr4} the equation for $R(X,Y,t)$:
\bea\label{eq:schr6}
\partial_t R(X,Y,t) & =  &  - \frac{i \hbar}{m} \sum_{j=1}^d \left( \nabla_{Y,j}^+ \nabla_{X,j}^-
+ \nabla_{X,j}^+ \nabla_{Y,j}^-  \right)R(X,Y,t)  \nonumber \\
 & & - \left( \frac{V_0}{2} \right)^2
\left[ g(0) - g(Y)  \right] R(X,Y,t) .
\eea
We write $\widetilde{R}(X,Y,s)$ for the Laplace transform of $R(X,Y,t)$ with respect to $t$. Taking the Laplace transform of \eqref{eq:schr6} we obtain:
\beq\label{eq:lat-lt9}
\frac{i\hbar}{m}  \sum_{j=1}^{d} \left[ \nabla_{Y,j}^+ \nabla_{X,j}^-
+ \nabla_{X,j}^+ \nabla_{Y,j}^- \right] {\widetilde{R}}(X,Y,s) + h(Y,s) {\widetilde{R}}(X,Y,s)
  =   {R}(X,Y,0),
  \eeq
where ${R}(X,Y,0) = \langle \rho_0 (x,x')\rangle$ and $h(Y,s) := s + \left( \frac{V_0}{2} \right)^2
\left[ g(0) - g(Y)  \right]$.

We next take the Fourier transform with respect to $X$,
\beq\label{eq:ft2}
\widehat{\widetilde{R}}(k,Y,s) := \sum_{X \in \Z^d} e^{i k \cdot X} \widetilde{R} (X,Y,s), ~~~k \in T^d.
\eeq
The Fourier transform of the differential operator term on the  left in \eqref{eq:lat-lt9} may be written as
\bea\label{eq:transform1}
2 \sum_{j=1}^{d} \left[ \left( e^{i k \cdot \hat{e}_j} - 1 \right) \widehat{\widetilde{R}} (k, Y + \hat{e}_j,s)
\right. &+& \left.
\left( e^{-i k \cdot \hat{e}_j} - 1 \right) \widehat{\widetilde{R}} (k, Y- \hat{e}_j,s)
\right.\nonumber \\
 &+& \left.
2 \left( 1 - \cos ( k \cdot \hat{e}_j ) \right) \widehat{\widetilde{R}} (k, Y,s) \right] .
\eea
By means of \eqref{eq:transform1}, the Fourier transform of \eqref{eq:lat-lt9} with respect to $X$ is
\bea\label{eq:lat-lt10}
\lefteqn{ \frac{ i \hbar}{m} \left( \sum_{j=1}^{d} \left[ \left( e^{i k \cdot \hat{e}_j} - 1 \right) \widehat{\widetilde{R}} (k, Y + \hat{e}_j,s)
+ \left( e^{-i k \cdot \hat{e}_j} - 1 \right) \widehat{\widetilde{R}} (k, Y- \hat{e}_j,s)  \right] \right)}  \nonumber \\
 & &  +  \frac{2 i \hbar}{m} \left( \sum_{j=1}^d \left( 1 - \cos ( k \cdot \hat{e}_j ) \right) \right) \widehat{\widetilde{R}} (k, Y,s)  +
 h(Y,s) \widehat{{\widetilde{R}}}(k,Y,s)    \nonumber \\
 &  & = \widehat{R}(k,Y,0).
\eea

In order to calculate the averaged mean square displacement, we compute $\frac{\partial^2}{\partial{k_m} \partial k_n} \widehat{\widetilde{R}}(k,0,s)|_{k=0}$ by
differentiating the equation \eqref{eq:lat-lt10} twice with respect to $k_j$ and then eliminating the $s$-dependent terms.
We first note that \eqref{eq:lat-lt10} evaluated at $k=0$ gives
\beq\label{eq:latticed9}
h(Y,s) \widehat{\widetilde{R}}(0,Y,s) = \widehat{{R}}(0,Y,0) .
\eeq
Let $c_1 := \frac{\hbar}{m}$ and we write $\partial_m := \frac{\partial}{\partial k_m}$.
The $k_m$-derivative of \eqref{eq:lat-lt10} at $k=0$ is
\beq\label{eq:lat-first-deriv1}
h(Y,s)\partial_m \widehat{\widetilde{R}}(0,Y,s) =  c_1 [  \widehat{\widetilde{R}}(0,Y + \hat{e}_m,s) -  \widehat{\widetilde{R}}(0,Y - \hat{e}_m ,s) ] + \partial_{m}\widehat{R}(0,Y,0).
\eeq
The mixed second partial derivative $\partial_{nm} : = \frac{\partial^2}{\partial k_n \partial k_m}$ of \eqref{eq:lat-lt10} at $k=0$
results in
\bea\label{eq:lat-second-deriv1}
\lefteqn{h(Y,s) {\partial_{nm}^2} \widehat{\widetilde{R}}(0,Y,s) - c_1[ \partial_m \widehat{\widetilde{R}}(0,Y + \hat{e}_n,s) - \partial_m \widehat{\widetilde{R}}(0,Y - \hat{e}_n,s)]} \nonumber \\
& & - c_1[ \partial_n \widehat{\widetilde{R}}(0,Y + \hat{e}_m,s) - \partial_n \widehat{\widetilde{R}}(0,Y - \hat{e}_m,s) \nonumber \\
 & & - i c_1 \delta_{mn} \left[ \widehat{\widetilde{R}}(0, Y+ \hat{e}_m,s) + \widehat{\widetilde{R}}(0,Y - \hat{e}_m,s) \right]  \nonumber \\
 & & + 2 i  c_1 \delta_{nm} \widehat{\widetilde{R}}(0,Y,s)
 = {\partial_{nm}^2} \widehat{R}(0, Y, 0).
 \eea
For the diagonal term $n=m$, with $\partial_m^2 := \frac{\partial^2}{\partial k_m^2}$, we obtain
\bea\label{eq:lat-second-deriv2}
\lefteqn{h(Y,s) {\partial_{m}^2} \widehat{\widetilde{R}}(0,0,s) - 2c_1[ \partial_m \widehat{\widetilde{R}}(0,Y + \hat{e}_m,s) - \partial_m \widehat{\widetilde{R}}(0,Y - \hat{e}_m,s)]} \nonumber \\
 & & - i c_1  \left[ \widehat{\widetilde{R}}(0, Y+ \hat{e}_m,s) + \widehat{\widetilde{R}}(0,Y - \hat{e}_m,s) \right]  \nonumber \\
 & & + 2 i  c_1  \widehat{\widetilde{R}}(0,Y,s)
 = {\partial_{m}^2} \widehat{R}(0, Y, 0).
 \eea

According to \eqref{eq:diffusion-condition1}, we need to extract the $s$-dependance of the terms on the right of \eqref{eq:lat-second-deriv2}
at $Y=0$:
\bea\label{eq:lat-second-deriv3}
h(0,s) {\partial_{m}^2} \widehat{\widetilde{R}}(0,0,s) & = &  2c_1[ \partial_m \widehat{\widetilde{R}}(0, \hat{e}_m,s) - \partial_m \widehat{\widetilde{R}}(0, - \hat{e}_m,s)] \nonumber \\
 & & + i c_1  \left[ \widehat{\widetilde{R}}(0,  \hat{e}_m,s) + \widehat{\widetilde{R}}(0, - \hat{e}_m,s) \right]  \nonumber \\
 & & - 2 i  c_1  \widehat{\widetilde{R}}(0,0,s) + {\partial_{m}^2} \widehat{R}(0, 0, 0).
 \eea

We use \eqref{eq:latticed9} and \eqref{eq:lat-first-deriv1}, and the evenness of $g$, to eliminate the factors of $\widehat{\widetilde{R}}$ depending on $s$ on the right of \eqref{eq:lat-second-deriv3}.
For the first term on the right in \eqref{eq:lat-second-deriv3}, we find:
\bea\label{eq:lat-second-deriv4}
[ \partial_m \widehat{\widetilde{R}}(0, \hat{e}_m,s) - \partial_m \widehat{\widetilde{R}}(0, - \hat{e}_m,s)] &=& c_1 [h(\hat{e}_m, s) h(2 \hat{e}_m, s)]^{-1} [  \widehat{{R}}(0, 2\hat{e}_m, 0) + \widehat{{R}}(0, - 2\hat{e}_m,0)] \nonumber \\
 & & - 2  c_1 [h(\hat{e}_m, s) h(0,s)]^{-1} \widehat{{R}}(0,0,0) \nonumber \\
 & & +  h( \hat{e}_m, s)^{-1} [ \partial_m \widehat{{R}}(0, \hat{e}_m, 0) - \partial_m \widehat{{R}}(0, - \hat{e}_m, 0)]
\eea
The second term may be written as
\beq\label{eq:lat-second-deriv5}
\left[ \widehat{\widetilde{R}}(0,  \hat{e}_m,s) + \widehat{\widetilde{R}}(0, - \hat{e}_m,s) \right] = h(\hat{e}_m, s)^{-1} \left[ \widehat{{R}}(0,  \hat{e}_m, 0) + \widehat{{R}}(0, - \hat{e}_m, 0) \right],
\eeq
and the third term,
\beq\label{eq:lat-second-deriv6}
\widehat{\widetilde{R}}(0, 0,s) = h(0,s)^{-1} \widehat{{R}}(0, 0,0).
\eeq

We use \eqref{eq:lat-second-deriv4}--\eqref{eq:lat-second-deriv6} in \eqref{eq:lat-second-deriv3}. This, combined with the facts that $h(0,s) =s$ and that the result must be real according to \eqref{eq:diffusion-condition1},
shows that
\beq\label{eq:diffusionZd}
- \Delta_k \widehat{\widetilde{R}}(0,0,s) =  \left( \frac{2 \hbar}{m} \right)^2 \left( \frac{1}{s^2} \right) \left\{ \sum_{m=1}^d \frac{1}{h(\hat{e}_m,s)} \right\} \widehat{R}(0,0,0) + \mathcal{O}\left( \frac{1}{s} \right) .
\eeq
We note that $h(\hat{e}_m,s) = s + (V_0/\hbar)^2 [g(0) - g(\hat{e}_m)]$. We assume that $g(0) \geq g(\hat{e}_m)$ for all $m=1, \ldots, d$. For example, if $g(x) = \tilde{g}(\|x\|)$, this condition is simply that $g(0) \geq g(1)$. If $g$ is strictly decreasing, this condition is satisfied.
Under this condition, the leading term of \eqref{eq:diffusionZd} is $\mathcal{O}(s^{-2})$ meaning the evolution is diffusive.
If, on the other hand, $g(0) = g(\hat{e}_m)$ for some $m$, then the leading term is $\mathcal{O}(s^{-3})$ meaning that the motion is ballistic. This also shows that the motion is also ballistic if $V_0 = 0$.

To explore this further, the first term on the right in \eqref{eq:diffusionZd} may be written as
$$
\sum_{m=1}^d ~\frac{C_d}{s^2 (s + \Gamma_m)} , ~~~~~~ \Gamma_m := (V_0/\hbar)^2 [g(0) - g(\hat{e}_m)],
$$
where $C_d := \left( \frac{2 \hbar}{m} \right)^2 \widehat{R}(0,0,0)$. We assume $\Gamma_m > 0$, for all $m$.
The inverse Laplace transform of this term is
\beq\label{eq:ilt1}
C_d \sum_{m=1}^d \left[ \frac{e^{- \Gamma_m t}}{\Gamma_m^2} + \frac{t}{\Gamma_m} - \frac{1}{\Gamma_m^2}   \right].
\eeq
If $\Gamma_m = 0$ for some $m$, then the inverse Laplace transform is
$$
\frac{C_d}{2} t^2,
$$
so the motion is ballistic.
For short times for which $t \Gamma_m \ll 1$, for all $m$, an expansion of the exponential in \eqref{eq:ilt1} yields the effective behavior
$$
\langle \|x\|^2 \rangle_{\rho_t} \sim  \left( \frac{d}{2} \right) t^2 . 
$$
This shows that for short times relative to $\Gamma_m$, the motion appears ballistic. If, on the other hand, $t  \Gamma_m \gg 1$, then the motion is diffusive and we obtain
$$
\langle \|x\|^2 \rangle_{\rho_t} \sim C_d \left( \sum_{m=1}^d \frac{1}{\Gamma_m} \right) t + {\rm constant} ,
$$
yielding the effective diffusion constant
\beq\label{eq:diff-cnst-latt1}
D(V_0,g) :=  \left( \frac{2 \hbar}{m} \right)^2 ~{\rm Tr} ( \rho_0 )  \left( \sum_{m=1}^d \frac{1}{\Gamma_m} \right)
\eeq
This result for the diffusion constant is reminiscent of the one-dimensional result of \cite{ovchinnikov-erikhman1}.
Their formula (52) for the averaged mean square displacement is our equation \eqref{eq:ilt1} and they find that the diffusion constant is proportional to $\Gamma^{-1}$, where $\Gamma$ is ${V_0^2}$, as in \eqref{eq:diff-cnst-latt1}.

\section{Appendix 1: SDE interpretation of the Schr\"odinger equation \eqref{eq:Schrod1}}\label{sec:sde1}

The Schr\"odinger equation \eqref{eq:Schrod1} and the quantum stochastic Liouville equation \eqref{eq:evolution1} are correctly interpreted as a stochastic differential equation (SDE) using the Stratonovich integral. For example,
for the Liouville equation, \eqref{eq:evolution1}, let $X_t$ denote the stochastic process $\rho(x,x',t)$
and let $L_0$ denote the deterministic Liouvillian
$$
(L_0f)(x,x') := \frac{\hbar^2}{2m} [ (\Delta_x f)(x,x') - (\Delta_{x'} f)(x,x')].
$$
We denote by $W_t$ the standard $d$-dimensional Brownian motion.
Then, we may write \eqref{eq:evolution1} as the SDE:
\beq\label{eq:sde1}
i \hbar d X_t = L_0 X_t ~dt +  X_t \circ (d W_t (x) - dW_t(x')),
\eeq
where $\circ$ denotes the Stratonovich integral.

With regard to the choice of the Stratonovich integral, we paraphrase from \cite[pgs. 62--63]{fischer-leschke-muller2}. The choice of the Stratonovich integral is quite natural on physical grounds.
The time change of a realistic physical system is governed by driving forces with a nonzero
correlation time. Theoretical models based on stochastic processes which are uncorrelated
in time may only be used successfully if the correlation time of the actual
driving forces is much smaller than any other time scale inherent to the system.
As in \cite{fischer-leschke-muller2}, we interpret equations \eqref{eq:Schrod1} and \eqref{eq:evolution1}
in the Stratonovich sense. Roughly speaking, this means that the white noise in \eqref{eq:sde1} may be replaced by colored noise for which the time correlation function $h_\nu$  is a smooth function converging weakly  to a delta function as $\nu \rightarrow 0$. The resulting regularized equation is averaged, and then the limit $\nu \rightarrow 0$ of the equation is taken. This procedure is used in the next section to compute a correlation function leading to equation \eqref{eq:schr4}. Justification for the procedure is given in a number of Wong-Zakai-like theorems, as in Horsthemke and Lefever \cite[pg.\ 101]{HorsthemkeLefever}, Karatzas and Shreve \cite[Chapter 5.2 D]{KaratzasShreve}, and Brzeiniak and Flandoli \cite{BrzFlandoli}. Wong-Zakai-like theorems guarantee that the solution $\rho_t^{(\nu)}$ of the regularized equation with colored noise parameterized by $\nu$ converges to the solution of the stochastic PDE with the Stratonovich interpretation.



\section{Appendix 2: An averaging result for Gaussian random variables}\label{sec:novikov1}

We consider a general situation where $V$ is a Gaussian random field with mean zero and  covariance function $C$ so that
\beq\label{eq:covar-gen1}
\langle V_\omega (x,t) V_\omega (y,s) \rangle = C((x,t), (y,s)).
\eeq
Let $R[V]$ be a functional of the Gaussian random variable $V$ with covariance function $C$ as in \eqref{eq:covar-gen1}. In this case, a result of Glimm-Jaffe \cite[Theorem 6.3.1]{glimmjaffe1} or Novikov \cite[section 2]{novikov1} states that
\beq\label{eq:average1}
\langle V(z,t) R[V] \rangle = \int_{\R^{d+1}} C((z,t),(y,s)) \left\langle \frac{\delta R[V]}{\delta V(y,s)} \right\rangle ~dy ~ds.
\eeq
In the white noise case, the covariance function is given by
\beq\label{eq:white-correl1}
C((x,t), (y,s)) = V_0^2 \delta (t-s) g(x-y),
\eeq
and for the colored noise case, we have a family of covariance functions
\beq\label{eq:colored-correl1}
C_nu ((x,t), (y,s)) = V_0^2 h_{\nu} (t-s) g(x-y),
\eeq
where $h_{\nu} (t)$ is a family of smooth functions with $h_{\nu}(t) \rightarrow \delta (t)$ in the distributional sense as $\nu \rightarrow 0$. We will assume that the support of $h_\nu \subset [ -\nu, \nu]$.

In keeping with the Stratonovich interpretation of the stochastic differential equation \eqref{eq:schr2} for the kernel $\rho (x,x',t)$, we will first compute the expectation \eqref{eq:average1} for colored noise with correlation function \eqref{eq:colored-correl1} and then take the limit $\nu \rightarrow 0$.

We write $V^{(\nu)}$ to denote colored noise and use \eqref{eq:average1} to first compute $\langle V^{(\nu)}(z,t) \rho (x',x,t) \rangle$, where $z$ denotes $x$ or $x'$. The functional $R[V]$ in \eqref{eq:average1} is $\rho [V^{(\nu)}](x',x,t)$. We write $\rho[V^{(\nu)}]$ to emphasize the dependence of $\rho$ on $V^{(\nu)}$.
According to \eqref{eq:average1}, we must compute the variational derivative of $\rho[V^{(\nu)}] (x',x,t)$ with respect to $V^{(\nu)}(y,s)$ and then take $s=t$ and finally $\nu \rightarrow 0$. We write the differential equation for the density matrix $\rho [V^{(\nu)}]$ in \eqref{eq:schr2} as
\bea\label{eq:ode-rho1}
\rho[V^{(\nu)}](x',x,t) & =  & - \frac{i \hbar}{2m} \int_0^t ( \Delta_{x'} - \Delta_x)\rho [V^{(\nu)}](x',x, \tau) d \tau  \nonumber \\
 & & + \left( \frac{i}{\hbar} \right) \int_0^t [ V^{(\nu)}(x', \tau) - V^{(\nu)}(x,\tau) ]  \rho [V^{(\nu)}] (x',x,\tau ) d\tau \nonumber \\
  & & +  \rho(x',x,0) ,
 \eea
where the initial density matrix $\rho (x',x,0)$ is independent of $V^{(\nu)}$.

Using the Liouvillian $L$ defined in \eqref{eq:generator1},
the variational derivative with respect to $V^{(\nu)}(y,s)$ may be computed from \eqref{eq:ode-rho1}. We note that the variation of the process at the time $\tau$ does not depend on the process at a later time $s > \tau$ so that
\beq\label{eq:history1}
\int_0^{s-\nu} ~L \left( \frac{\delta \rho [V^{(\nu)}](x',x,\tau)}{\delta V^{(\nu)}(y,s)} \right) ~d \tau = 0 .
\eeq
As a consequence, we obtain from \eqref{eq:ode-rho1}:
\bea\label{eq:variational1}
\frac{\delta \rho [V^{(\nu)}](x',x,t)}{\delta V^{(\nu)}(y,s)} &=& \frac{i}{ \hbar} \int_{s-\nu}^t ~L \left( \frac{\delta \rho [V^{(\nu)}](x',x,\tau)}{\delta V^{(\nu)}(y,s)} \right) ~d \tau \nonumber \\
 & & + \frac{i}{\hbar} \int_0^t \left\{ \frac{[ \delta V^{(\nu)}(x',\tau) - \delta V^{(\nu)}(x, \tau)]}{ \delta V^{(\nu)}(y,s) } \right\} ~
    \rho [V^{(\nu)}] (x',x,\tau ) d\tau . \nonumber \\
     & &
\eea
For the argument of the second term on the right in \eqref{eq:variational1}
we compute
\beq\label{eq:variational2}
\left\{ \frac{[ \delta V^{(\nu)}(x',\tau) - \delta V^{(\nu)}(x, \tau)]}{ \delta V^{(\nu)}(y,s) } \right\} =  h_{\nu} (\tau - s) [ \delta (x'-y) - \delta (x-y) ] .
\eeq
Taking the limit $s \rightarrow t$, with $s \leq t$, we obtain from \eqref{eq:variational1}--\eqref{eq:variational2}:
\bea\label{eq:variational3}
\lim_{s \rightarrow t} \frac{\delta \rho [V](x',x,t)}{\delta V(y,s)} &=& \frac{i}{ \hbar} \int_{t-\nu}^t ~L \left( \frac{\delta \rho [V^{(\nu)}](x',x,\tau)}{\delta V^{(\nu)}(y,s)} \right) ~d \tau \nonumber \\
  & & + \frac{i }{\hbar} \int_0^t h_\nu (t-\tau) [ \delta (x' -y) - \delta (x-y)]
\rho[V](x', x, \tau) ~d \tau . \nonumber \\
\eea
We now take $\nu \rightarrow 0$. The limit of the first term on the right in \eqref{eq:variational3} vanishes and the limit of the second term may be evaluated using $\delta(\tau) = \frac{d}{d \tau} H(\tau)$, where $H(\tau)$ is the Heaviside function with $H(0) = \frac{1}{2}$.
Inserting this result \eqref{eq:variational3} into \eqref{eq:average1}, we obtain
\bea\label{eq:average-final1}
\langle V(z,t) \rho[V](x',x,t) \rangle  &=& \frac{iV_0^2}{2\hbar} \int_{\R^d} g(z-y) [ \delta (x' -y) - \delta (x-y)]
\langle \rho[V](x', x, t) \rangle ~dy \nonumber \\
 &=& \frac{i V_0^2}{2\hbar} [ g(z-x') - g (z-x)] \langle \rho[V](x', x, t) \rangle  .
\eea
This establishes the result needed in the derivation of \eqref{eq:averaging-result1}.

\begin{acknowledgement}
PDH thanks P.\ M\"uller for several discussions on his work \cite{muller1} and on \cite{fischer-leschke-muller, fischer-leschke-muller2}, S.\ De Bi\`evre for discussions on classical systems, and J.\ Marzuola for discussions on stochastic PDEs. PDH was partially supported by NSF DMS 11-03104, KK was partially supported by NSF DMS-1106770, CAREER DMS-1254791 and a Simons Sabbatical Fellowship, SO was partially supported by the ANR-15-CE40-0020-01 grant LSD, and JS was partially supported by NSF DMS-1500386, while some of this work was done.

\end{acknowledgement}

\end{document}